\documentclass[preprint,prc,letterpaper,11pt,twoside,tightenlines,nofootinbib,showpacs]{revtex4-1}
\usepackage{amsmath,amssymb,bm}
\usepackage{color}
\usepackage{graphicx}
\usepackage{epstopdf}

\newcommand{\dd}{\mathrm{d}}
\newcommand{\MeV}{\mathrm{MeV}}
\newcommand{\GeV}{\mathrm{GeV}}
\newcommand{\TeV}{\mathrm{TeV}}

\newcommand{\bvec}[1]{\ensuremath{\boldsymbol{#1}}}

\begin{document}

\preprint{APS/123-QED}

\title{Dileptons from correlated D- and $\overline{\text{D}}$-meson decays in the invariant mass range of the QGP thermal radiation using the UrQMD hybrid model}

\author{Thomas Lang$^{1,2}$} \author{Hendrik van Hees$^{1,2}$} \author{Jan Steinheimer$^{3}$}
\author{Marcus Bleicher$^{1,2}$}
\affiliation{
  $^{1}\,$Frankfurt Institute for Advanced Studies
  (FIAS),Ruth-Moufang-Str. 1, 60438 Frankfurt am Main, Germany }
\affiliation{
  $^{2}\,$Institut f\"ur Theoretische Physik, Johann Wolfgang
  Goethe-Universit\"at, Max-von-Laue-Str. 1, 60438 Frankfurt am Main,
  Germany }
\affiliation{
  $^{3}\,$Lawrence Berkeley National Laboratory, 1 Cyclotron Road, Berkeley, CA 94720, USA
}

\date{\today}

\begin{abstract}
  Relativistic heavy-ion collisions produce a hot and dense thermalized
  medium, that is expected to emit thermal radiation in form of
  dileptons. These dileptons are not affected by the strong force and
  are therefore a clean probe for the possible creation of a Quark Gluon
  Plasma (QGP). However, electroweak decays of open-charm mesons are
  expected to induce a substantial background in the invariant mass region
  between the $\phi$ and $J/\Psi$ peak ($1 \; \GeV \lesssim M_{\ell^+
    \ell^-} \lesssim 3 \; \GeV$) of the thermal QGP radiation.  To
  evaluate this background radiation we apply a Langevin approach for
  the transport of charm quarks in the UrQMD (hydrodynamics + Boltzmann)
  hybrid model. Due to the inclusion of event-by-event fluctuations and
  a full (3+1)-dimensional hydrodynamic evolution, the UrQMD hybrid
  approach provides a more realistic model for the evolution of the
  matter produced in heavy ion collisions as compared to simple
  homogeneous fireball expansions usually employed before. As drag and
  diffusion coefficients we use a resonance approach for elastic
  heavy-quark scattering and assume a decoupling temperature of the
  charm quarks from the hot medium of $130 \; \MeV$. For the
  hadronization of the charm quarks we employ a coalescence approach at
  the decoupling temperature from the medium. In this letter we present
  our calculations of the D-meson correlations and the invariant mass
  spectra of the dilepton decays in heavy-ion collisions at FAIR, RHIC,
  and LHC energies using different interaction scenarios.
\end{abstract}

\maketitle


\section{Introduction}
\label{sec:Langevin}
\markright{Thomas Lang, Hendrik van Hees, Jan Steinheimer, and Marcus
  Bleicher. Medium modification of charm quarks at FAIR in a Langevin
  approach.}

In the spirit to explore the possible existence of a phase of matter
with quarks and gluons as degrees of freedom, the Quark Gluon Plasma (QGP), various
attempts have been made to propose observables for its verification
\cite{Adams:2005dq,Adcox:2004mh,Muller:2012zq}. 

One particularly interesting probe for the exploration of the QGP are 
charm quarks. Charm quarks are produced in the hard primary parton-parton
collisions in the pre-equilibrium phase of the matter evolution. Since
the quantum number charm is conserved in the strong interaction, 
charm quarks are produced as charm-anti-charm pairs and are emitted dominantly
back-to-back due to momentum conservation. In the following evolution of
the (locally thermalized) medium, the charm quarks traverse the QGP and finally hadronize 
to D/$\bar{\text{D}}$-mesons when the system cools down. These
D/$\bar{\text{D}}$-mesons decay to electrons and positrons and can be measured.  The angular
correlations between the decaying D/$\bar{\text{D}}$-mesons lead to correlated $e^+e^-$-pairs 
which populate the invariant-mass spectra. 
Depending on the strength of the medium interaction of the D/$\bar{\text{D}}$-mesons and their
momentum change during the hadronization process a different strength in 
the angular correlations have already been explored in \cite{Zhu:2006er,Schweda:2006qc} 
using a Langevin approach on a Bjorken's hydrodynamics background. 
In this study a substantial modification of the correlations distributions 
in the partonic phase was found. The modification in the hadronic phase, 
however, seems to be negligible. \\

Apart from learning more about D-meson interactions in the medium, 
such calculations can also be utilized to learn more about thermal QGP radiation. 
Thermal dilepton radiation of the
hot and dense matter allows to draw conclusions on the matters nature, i.e., to
determine its temperature and the in-medium properties of the
electromagnetic current correlation function, which is closely related
to the spectral properties of the light vector mesons, $\rho$, $\omega$,
and $\phi$, in the low-mass region, $M_{\ell^+ \ell^-} \lesssim 1 \;
\GeV$
\cite{Rapp:2000pe,Gallmeister:1998ja,Shuryak:1996gc}. Unfortunately
though, background radiation complicates these measurements. By far, the
most important contribution to this background radiation in the
invariant-mass range between the $\phi$ and $J/\Psi$ peak of
approximately $1\;\text{GeV}$ to $3\;\text{GeV}$ is the dilepton
radiation originating from open-charm decays, where it competes with the
thermal radiation from the partonic phase of the fireball
evolution. This background yield is sensitive to the energy loss of
charm quarks in the medium. Thus, the knowledge of charm quark and 
D-meson interactions in the hot medium provides us with insights to the 
thermal QGP radiation. \\

In this paper we explore the D-meson correlations and the invariant mass
spectra of D-meson dileptons at FAIR (Au+Au at a beam energy of
$E_{\text{lab}}=25\,\text{GeV}$), RHIC (Au+Au at $\sqrt{s_{NN}}=200\;\text{GeV}$) and
LHC (Pb+Pb at $\sqrt{s_{NN}}=2.76\;\text{TeV}$).  For this purpose we use
a hybrid model, consisting of the Ultra-relativistic Quantum Molecular
Dynamics (UrQMD) model \cite{Bass:1998ca,Bleicher:1999xi} and a full
(3+1)-dimensional ideal hydrodynamical model
\cite{Rischke:1995ir,Rischke:1995mt} to simulate the bulk medium. The
heavy-quark propagation in the medium is described by a relativistic
Langevin approach.

Similar quark-propagation studies have recently been performed in a
thermal fireball model with a combined coalescence-fragmentation
approach from Rapp and Hees 
\cite{vanHees:2007me,vanHees:2007mf,Greco:2007sz,vanHees:2008gj,Rapp:2008fv,
  Rapp:2008qc,Rapp:2009my}, in an ideal hydrodynamics model with a
lattice-QCD EoS \cite{He:2012df,He:2012xz}, in a model from Kolb and
Heinz \cite{Aichelin:2012ww}, in the BAMPS model
\cite{Uphoff:2011ad,Uphoff:2012gb}, the MARTINI model
\cite{Young:2011ug} as well as in further studies and model comparisons
\cite{Moore:2004tg,Vitev:2007jj,Gossiaux:2010yx,Gossiaux:2011ea,Gossiaux:2012th}.
Previous studies, that focus on the correlation of D-mesons and/or
invariant mass spectra in transport models, can be found in
\cite{Zhu:2006er,Schweda:2006qc,Bratkovskaya:2012st,Linnyk:2012pu,Nahrgang:2013saa}.

The UrQMD hybrid model provides a realistic and well established background, including
event-by-event fluctuations and has been shown to very well describe
many collective properties of the hot and dense medium created in
relativistic heavy-ion collisions.

\section{Description of the model}

The UrQMD hybrid model \cite{Petersen:2008dd} combines the advantages of
transport theory and (ideal) fluid dynamics. The initial conditions are 
generated by the UrQMD model
\cite{Bass:1999tu,Dumitru:1999sf,Bleicher:1999xi}, followed by a full (3+1)-dimensional ideal
fluid dynamical evolution, including the explicit propagation of the
baryon current. After a Cooper-Frye transition back to the transport
description, the freeze out of the system is treated dynamically within
the UrQMD hadron cascade. The hybrid model has been successfully applied to
describe particle yields and transverse dynamics from AGS to LHC
energies
\cite{Petersen:2008dd,Steinheimer:2007iy,Steinheimer:2009nn,Petersen:2010cw,Petersen:2011sb}
and is therefore a reliable model for the flowing background medium.

The equation of state includes quark and
gluonic degrees of freedom coupled to a hadronic parity-doublet model
\cite{Steinheimer:2011ea}. It has a smooth crossover at low baryon
densities between an interacting hadronic system and a quark gluon
plasma and a first order phase transition at higher densities. 
The thermal properties of the EoS are in agreement with lattice
QCD results at vanishing baryon density. 

For the present study of the charm quark dynamics in the expanding medium of light quarks 
we employ the well-known stochastic Langevin equation, suitable for numerical simulations 
\cite{Svet88,MS97,Moore:2004tg,vanHees:2005wb,vanHees:2007me,Gossiaux:2008jv,He:2011yi}. 
Such a Langevin process reads
\begin{equation}
\label{lang.1}
\dd x_j = \frac{p_j}{E} \dd t, \qquad \dd p_j = -\Gamma p_j \dd t + \sqrt{\dd t} C_{jk} \rho_k.
\end{equation}
Here $E=\sqrt{m^2+\bvec{p}^2}$, and $\Gamma$ is the drag or friction
coefficient. The covariance matrix, $C_{jk}$, of the fluctuating force
is related with the diffusion coefficients. Both coefficients are
dependent on $(t,\bvec{x},\bvec{p})$ and are defined in the (local) rest
frame of the fluid. The $\rho_k$ are Gaussian-normal distributed random
variables. 
For details, the reader is referred to \cite{Rapp:2009my,Lang:2012cx}.

\begin{figure}[h!]
\begin{minipage}[b]{0.45\textwidth}
\includegraphics[width=1\textwidth]{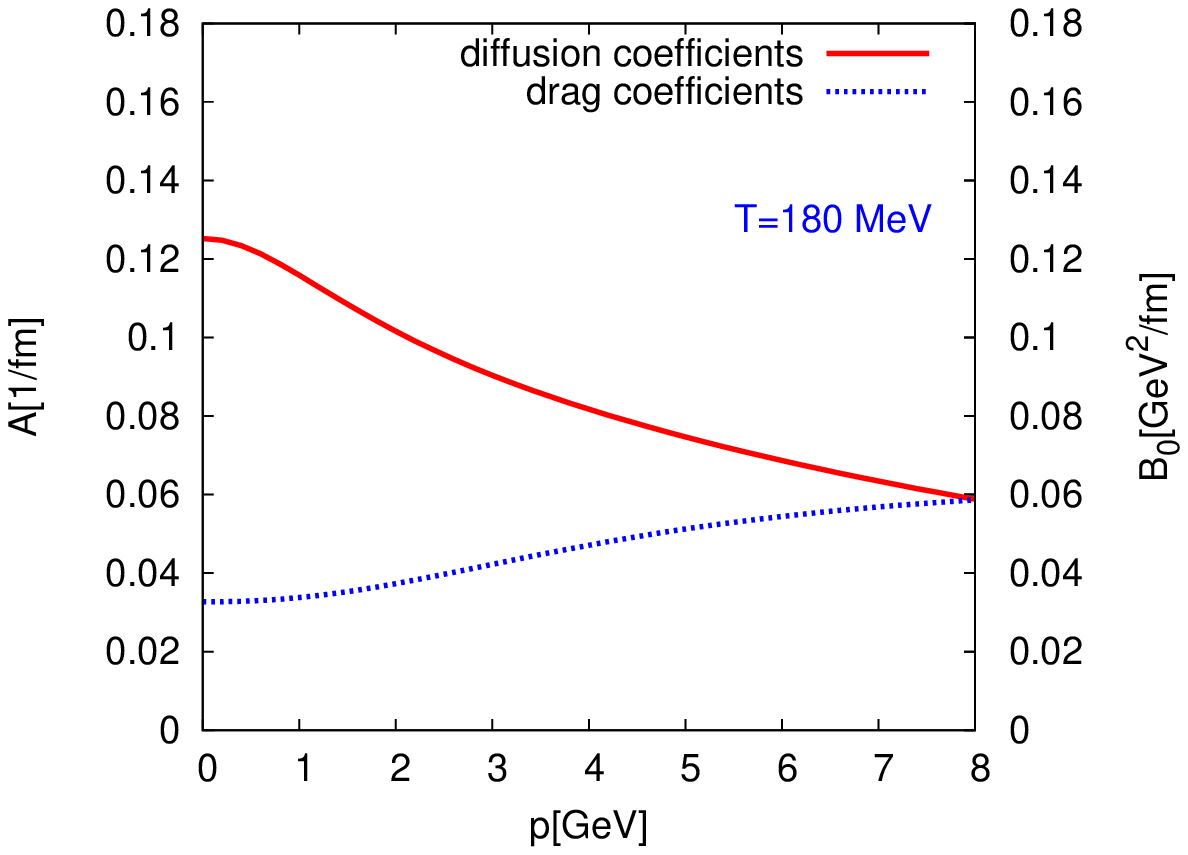}
\end{minipage}
\hspace{5mm}
\begin{minipage}[b]{0.45\textwidth}
\includegraphics[width=1\textwidth]{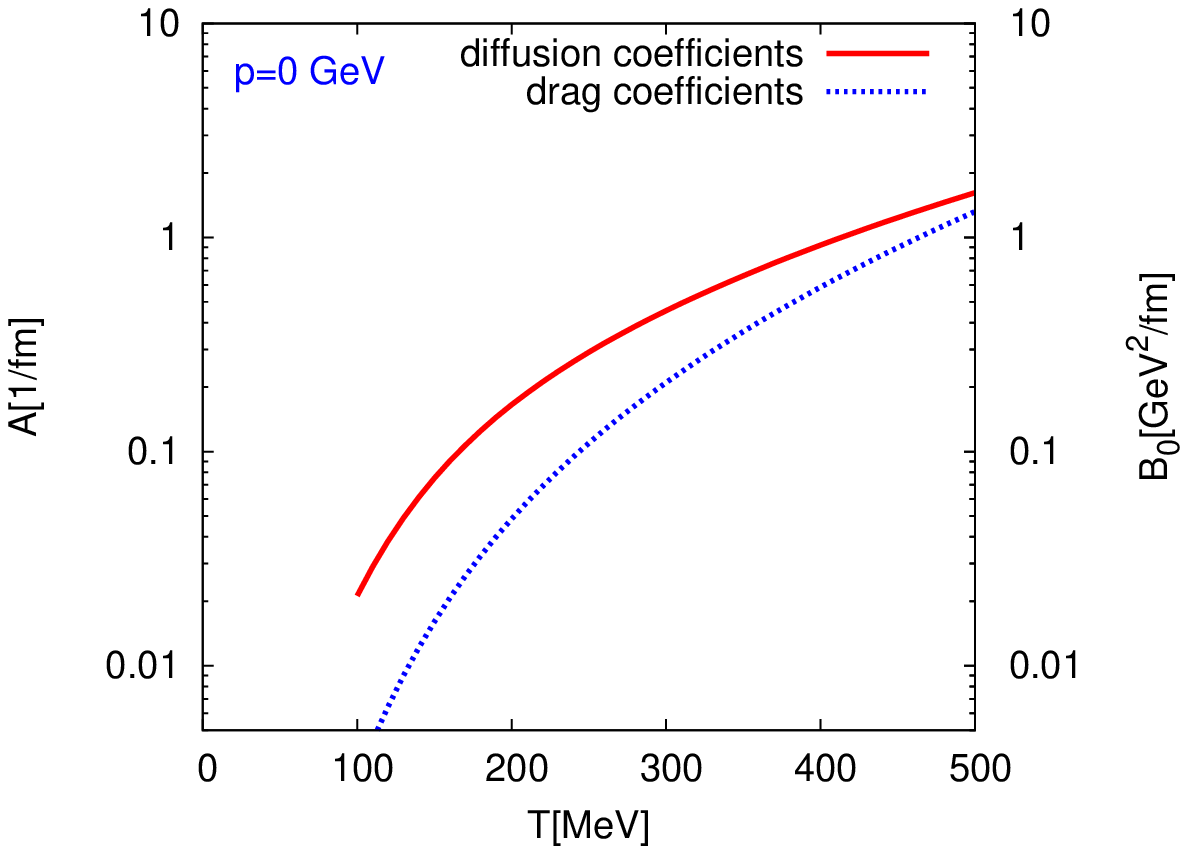}
\end{minipage}
\caption[Coefficients in the resonance model]
{(Color online) Drag and diffusion coefficients in the resonance
  model for charm quarks. 
  Left: The plot
  shows the dependence of the coefficients on the three-momentum
  $|\vec{p}|$ at a fixed temperature of $T=180\,\text{MeV}$. 
  Right: The plot
  shows the dependence of the coefficients on the temperature at a fixed
  three-momentum $|\vec{p}|=0$. }
\label{Coeffp}
\end{figure}

The drag and diffusion coefficients for the heavy-quark propagation
within this framework are taken from a resonance approach
\cite{vanHees:2005wb,Lang:2012cx}. It is a non-perturbative approach, where the
existence of D-mesons and B-mesons in the QGP phase is assumed. 
The drag and diffusion coefficients obtained from this approach are shown in 
Figure \ref{Coeffp} (left) as function of
the three-momentum $|\vec{p}|$ at $T=180\,\text{MeV}$ and in Figure 
\ref{Coeffp} (right) as function of the temperature at a fixed three-momentum of
$|\vec{p}|=0$. 
 \\

The initial production of charm quarks in our framework is based on a
time resolved ``Glauber'' approach \cite{Bialas:1976ed,Glauber:1987bb,Spieles:1999kp,Miller:2007ri}, 
i.e.\ we perform first a UrQMD run excluding
interactions between the colliding nuclei and save the nucleon-nucleon
collision space-time coordinates.  These coordinates are used in a
second, full UrQMD run as (possible) production space-time coordinates for the
charm quarks.

For the initially produced
charm quarks at FAIR, RHIC and LHC energies we utilize different models: 
\begin{itemize}
 \item 
For collisions at $E_{\text{lab}}=25\,\text{AGeV}$ we use the D-meson transverse mass spectrum from HSD
calculations \cite{Cassing:2000vx} fitted by  
\begin{equation}
\frac{\dd N}{\dd p_T} =\frac{C}{\left(1+A_1\cdot p_T^2\right)^{A_2}},
\end{equation}
with the coefficients $A_1=\,0.870/\text{GeV}^2$ and $A_2=\,3.062$. 
$C$ is an arbitrary normalization constant with the unit $1/\text{GeV}$. 
This function is then used for the initial charm
quark production. 
\item
At RHIC
($\sqrt {s_{NN}}=200\; \GeV$) we use
\begin{equation}
\frac{\dd N}{\dd p_T}=\frac{C\cdot\left(1 + A_1\cdot p_T^2\right)^2p_T}{\left(1+A_2\cdot p_T^2\right)^{A_3}},
\end{equation}
with $A_1=2.0/\text{GeV}^2$, $A_2=0.1471/\text{GeV}^2$, $A_3=21.0$. 
$C$ is an arbitrary normalization constant with the unit $1/\text{GeV}^2$. 
This distribution is taken
from \cite{vanHees:2005wb,vanHees:2007me}. It is obtained by using tuned
c-quark spectra from PYTHIA.  Their pertinent semileptonic
single-electron decay spectra account for pp and dAu measurements by the
STAR collaboration up to $p_T=4\,\text{GeV}$. The missing part at higher
$p_T$ is then supplemented by B-meson contributions. 
\item
At LHC ($\sqrt {s_{NN}}=2.76\; \TeV$) the initial distribution is
obtained from a fit to PYTHIA calculations. The fitting function we use
is
\begin{equation}
\frac{\dd N}{\dd p_T} =\frac{Cp_T}{\left(1+A_1\cdot p_T^2\right)^{A_2}}
\end{equation}
with the coefficients $A_1=0.379/\text{GeV}^2$ and $A_2=\,5.881$. 
$C$ again is an arbitrary normalization constant with the unit $1/\text{GeV}^2$. 
In all cases we neglect gluon radiation on the outgoing charm quarks and 
assume a full back-to-back emission of the $c\bar{c}$ pair. 
\end{itemize}

Starting with these charm-quark distributions as initial
conditions we perform in the hydrodynamic state an 
Ito post-point time-step of the Langevin simulation, employing the 
local hydro's cell velocities and cell temperatures for the calculation of 
the momentum transfer to the heavy quarks. 

We include a 
hadronization mechanism for charm quarks into D-Mesons, via the use of a quark-coalescence
mechanism (see \cite{Lang:2012yf,Lang:2012cx,Lang:2012vv,Lang:2013cca} for details) 
when the decoupling temperature is reached. 

This approach has already been successfully applied to describe 
experimental measurements of the nuclear modification factor and the elliptic flow 
at RHIC and LHC energies \cite{Lang:2012cx,Lang:2012yf,Lang:2012vv}. Moreover it has been used to predict 
the medium modification of D-mesons at FAIR energies \cite{Lang:2013cca}.

\section{The correlation-angle distribution}

\begin{figure}[h!]
\center
\includegraphics[width=0.6\textwidth]{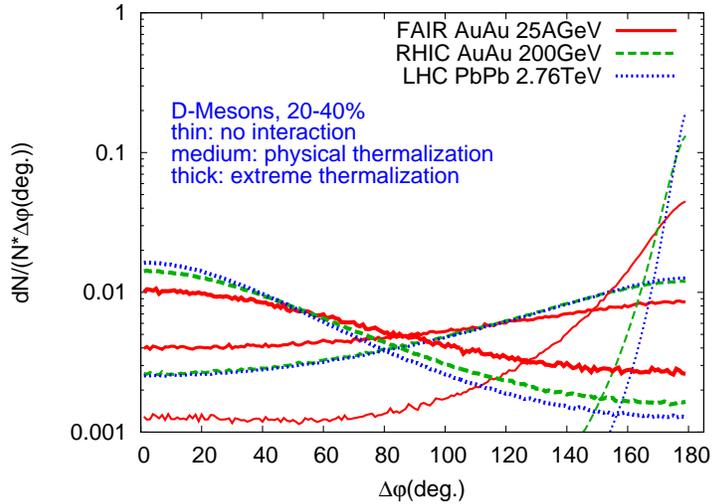}
\caption{(Color online) Relative azimuth correlations of D-mesons at FAIR, RHIC
  and LHC energies in the centrality range of 20-40\%. Three different
  scenarios are shown: no interaction with the medium, physical drag and
  diffusion coefficients and extreme thermalization with 20-fold higher
  coefficients. The angular
  modification for the case of no interaction with the medium is due to the coalescence
  mechanism. The yields are normalized to one. }
\label{azicorrelations}
\end{figure}

Let us start with our results for the azimuth angular correlation, 
followed by the full angular correlation. 
We have performed our calculations for FAIR, RHIC and LHC energies in
the centrality range of $\sigma /\sigma_{\text{tot}}=\,$20\%-40\%. 
Our results for the relative azimuth correlation of the decoupled D-mesons, 
that means the azimuth angle of the momenta between quark pairs 
emitted back-to-back, 
are shown in Fig.\ \ref{azicorrelations}. 

The following observations can be made: 

No interaction: The charm quarks do not interact with the medium. 
  The modification of the initial back-to-back correlation is due to the coalescence 
  mechanism only. We observe that the coalescence alone leads to a considerable 
  modification of the D$\bar{\text{D}}$ angular distribution. 
  This modification grows larger for lower energies since the charm
  momenta are lower and the modification due to the coalescence with 
  the light quarks is relatively more important. 

Physical scenario: We apply the full calculation plus coalescence,
  which has been successful in describing $R_{AA}$ and $v_2$ of
  non-photonic electrons at RHIC as well as D-mesons at the LHC \cite{Lang:2012cx,Lang:2012yf,Lang:2012vv}. 
  Here the angular distribution flattens and virtually
  all charm quarks interact considerably with the hot medium. The
  distribution for charm quarks at RHIC and LHC energies does hardly
  differ while the modification at FAIR energies is somewhat larger due
  to the slower medium evolution and the stronger modification due to
  the coalescence mechanism. This finding is in agreement with the calculations of \cite{Zhu:2006er}. 

Extreme heavy-quark thermalization: For this scenario the drag and
  diffusion coefficients are multiplied by a factor of 20. Therefore the
  charm quarks become nearly completely thermalized with the bulk
  medium. The charm quarks tend to adopt angular correlations diametrically opposed 
to the initial correlations. This is different from the calculated ``flat'' correlation distribution 
for full thermalization in \cite{Zhu:2006er}. This difference is due to the radial and elliptic flow 
which has been neglected in the calculation of \cite{Zhu:2006er}, but is included in our full (3+1)-dimensional hydrodynamic evolution. 
Charm and anti-charm-quarks which are initially produced back-to-back are dragged with the expanding medium and their 
angles align. \\

\begin{figure}[h!]
\center
\includegraphics[width=0.6\textwidth]{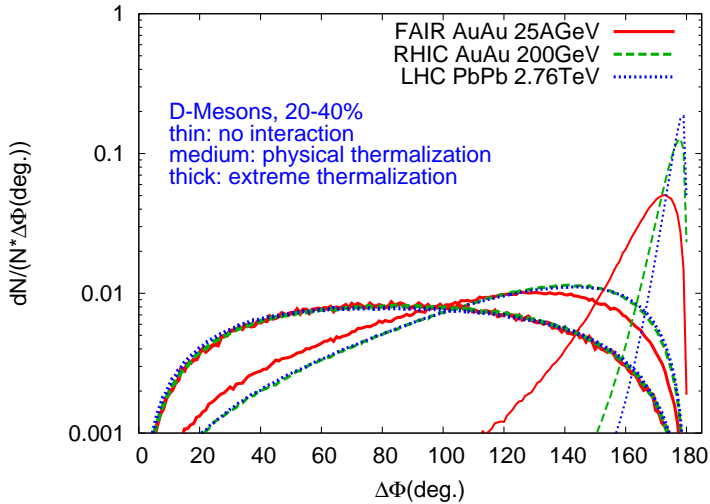}
\caption{(Color online) Angular correlations of D-mesons at FAIR, RHIC
  and LHC energies in the centrality range of 20-40\%. Three different
  scenarios are shown: no interaction with the medium, physical drag and
  diffusion coefficients and extreme thermalization with 20-fold higher
  coefficients.  The correlation gets flatter for lower collision
  energies and larger drag and diffusion coefficients. The angular
  modification for the case of no interaction is due to the coalescence
  mechanism. The yields are normalized to one. }
\label{correlations}
\end{figure}

In the following we will explore the full angular correlations of D-mesons defined by 
\begin{equation}
\Delta\phi=\cos^{-1}\left(\vec{u}\cdot\vec{v}\right)/\left(|\vec{u}|\cdot|\vec{v}|\right) . 
\end{equation}
Our results are shown in Fig.\ \ref{correlations}.

Here, for random angular correlations the distribution corresponds to the 
geometrical $\sin{\Delta\Phi}$ dependence. As already seen in case of the azimuthal correlations, 
the correlations are moved to smaller angles for higher drag and diffusion coefficients. 
For the extreme thermalization scenario the D-mesons are preferably correlated with small angles 
due to the substantial radial and elliptic flow, as already seen in Fig.\ \ref{azicorrelations}. 

\section{The invariant mass spectra of electrons from D-meson decays}

As described in the introduction, electrons/positrons from D$\bar{\text{D}}$-meson decays are the main 
background contribution for thermal QGP radiation in the invariant mass region 
of $1$ to $3\,\text{GeV}$. Therefore, these invariant mass spectra will 
be explored for FAIR, RHIC and LHC energies in the following. Finally, 
the invariant mass spectrum at RHIC energy will be compared to available data. \\

\begin{figure}[h]
\center
\includegraphics[width=0.6\textwidth]{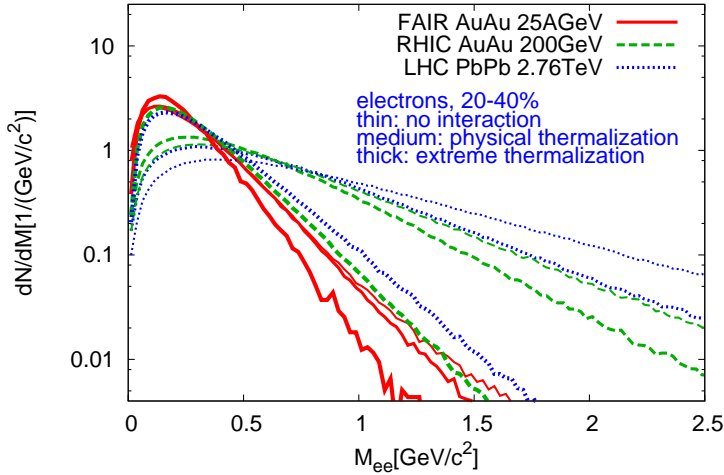}
\caption{(Color online) Invariant mass spectra of electrons from D-meson
  decays at FAIR, RHIC, and LHC energies in the of 20-40\% centrality
  class. Three different scenarios are shown: no interaction with the
  medium, physical drag and diffusion coefficients and extreme
  thermalization with 20-fold higher coefficients.  The spectra are
  harder for higher collision energies. Thermalization leads to
  considerably softer invariant mass spectra. The yields are normalized to one. }
\label{invariant}
\end{figure}

To obtain these invariant mass spectra, we have described the decay of the (still correlated) D- and
$\overline{\text{D}}$-mesons to $\text{e}^+ \text{e}^-$ pairs using
PYTHIA. Our results
are shown in Fig.\ \ref{invariant}.

Again the results for the three different energies assuming no
thermalization, normal thermalization and extreme thermalization are
shown. As expected the invariant mass spectra are softer for lower
collision energies. At all energies the different thermalization
scenarios lead to considerably different results. High thermalization
leads to a strong suppression of high momentum particles, and the
spectra become softer. In contrast to our calculations for the angular correlation 
we see differing results for the different collision energies 
also when applying the full thermalization scenario. 
The reason are the higher momenta of the D-mesons at higher collision
energies which translate to harder decay spectra.

\begin{figure}[h!]
\center
\includegraphics[width=0.6\textwidth]{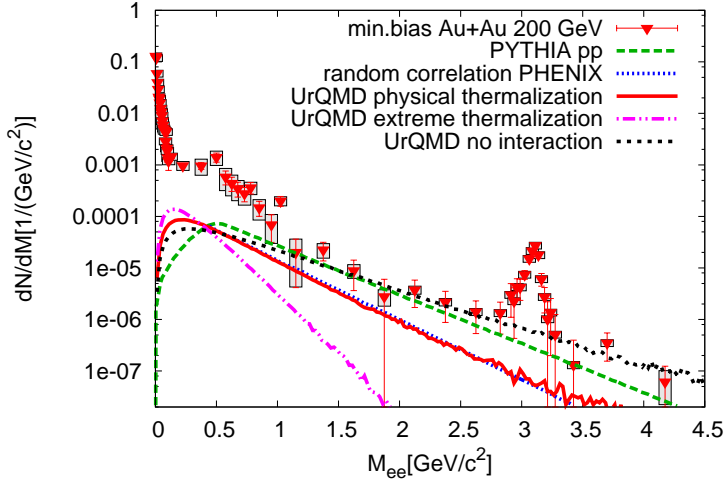}
\caption{(Color online) Invariant mass spectra of electrons in Au+Au
  collisions at $\sqrt{s}_{NN}=200\,\text{GeV}$ in the centrality range
  of 0-92\% (min. bias). A rapidity cut of $|y|<0.35$ and a momentum cut
  of $p_T^e=0.2\text{GeV/c}$ are applied. The dilepton data points are
  taken from a PHENIX measurement \cite{Adare:2009qk}. They are compared
  to calculations of electrons from D-meson decays. Also the pp
  calculation in PYTHIA and the random correlation calculation are taken
  from PHENIX \cite{Adare:2009qk}. The UrQMD calculations show three
  different scenarios: no interaction with the medium, physical drag and
  diffusion coefficients and extreme thermalization with 20 times higher
  coefficients. The difference between our physical scenario and the
  measured dilepton decays might be due to thermal radiation from the
  medium. The yields are normalized to the PYTHIA pp yield taken from \cite{Adare:2009qk}.}
\label{RHICdata}
\end{figure}

Finally we compare our calculation to an experimentally measured
invariant-mass spectrum from the PHENIX experiment at RHIC.  Here we
have performed a minimum bias calculation ($\sigma
/\sigma_{\text{tot}}=0\%$-$92\%$) and applied the appropriate
experimental cuts. The result is shown in Fig.\ \ref{RHICdata}.

Here especially the $M_{\ell^+ \ell^-}$ range between the $\phi$- and
$J/\Psi$ peak of $\sim 1$ to $3\,\text{GeV}$ is of large interest because
thermal dilepton radiation from the hot and dense matter is expected to
dominate among all other thermal sources in this range. Our UrQMD
calculation excluding interactions of the charm quarks with the medium
approximately matches the measured dilepton radiation in the range of
$1$ to $3\;\text{GeV}$. If we include interactions of the charm quarks
with the medium in the calculation (physical thermalization scenario) 
our result well underestimates the experimental
data. This difference between the dilepton contribution of D-meson
decays and the experimental dilepton measurements might originate from 
thermal radiation and might therefore be a QGP signal.  It increases
significantly in the dilepton mass range from $1$ to $3 \; \text{GeV}$.
For completion we also included our calculation for extreme thermalization.
Here high invariant masses are strongly suppressed and the difference
between our calculation and the measured data is substantial in the
invariant mass range of interest. \\
 
The difference of our calculation excluding interactions and the pp
calculation in PYTHIA is, besides the hadronization mechanism, due to
the different initial charm-quark distributions used. The random
correlation that is used by the PHENIX collaboration assumes a flat
angular correlation in contrast to the results of the calculation (cf.\ Fig.\
\ref{correlations}).

\section{Summary}

In this paper we have calculated the angular correlations of D-mesons
and the invariant-mass spectra of $\mathrm{e}^+ \mathrm{e}^-$ pairs from
correlated D- and $\overline{\text{D}}$-meson decays using a Langevin
simulation within the UrQMD hybrid model for the bulk-medium evolution
at FAIR, RHIC and LHC energies.  For these calculations we have assumed
a scenario without interactions of charm quarks with the medium, a
physical scenario with realistic drag and diffusion coefficients
(providing a satisfactory description of data on the $R_{AA}$ and $v_2$
of non-photonic electrons at RHIC and D-mesons at LHC) and a scenario of
extreme thermalization with strongly enhanced drag and diffusion
coefficients.

The azimuthal correlation distributions show a flattening of the 
distribution function for higher drag and diffusion coefficients as 
predicted in an earlier calculation \cite{Zhu:2006er}. 
For an extreme thermalization, however, the correlation function tends 
to small angles do to the radial and elliptic flow developing in 
our hydrodynamic calculation. 
This effect can also be observed when exploring the full angular 
correlation distribution. 

Subsequently, we
have studied the invariant mass spectra of the electrons from D-meson
decays and find softer spectra for low collision energies. Moreover the
spectra show a considerably softer shape in case of stronger
heavy-flavour-medium interactions.

Finally we have compared our calculations to experimental data from the
PHENIX experiment.  Here we find that our physical scenario cannot 
saturate the dilepton radiation in the dilepton invariant mass range
between $1$ and $3 \;\text{GeV}$ as a single source. We deduce that
the difference between the experimental measurement and our calculation
might be due to thermal radiation of the hot medium and might therefore
originate from thermal QGP radiation. 

We conclude, that a measurement of the angular correlation function 
would allow to check the validity of our model approach for D-meson 
correlations and therefore directly lead to a more definite estimation of the 
QGP background radiation. 

\section{ACKNOWLEDGMENTS}

We are grateful to the Center for Scientific Computing (CSC) and the
LOEWE-CSC at Frankfurt for providing computing resources. T.~Lang
gratefully acknowledges support from the Helmholtz Research School on
Quark Matter Studies. This work is supported by the Hessian LOEWE
initiative through the Helmholtz International Center for FAIR (HIC for
FAIR). J.~S. acknowledges a Feodor Lynen fellowship of the Alexander von
Humboldt foundation.  This work is supported by the Office of Nuclear
Physics in the US Department of Energy's Office of Science under
Contract No. DE-AC02-05CH11231, the GSI Helmholtzzentrum and the
Bundesministerium f{\"ur} Bildung und Forschung (BMBF) grant
No. 06FY7083. 

\bibliography{bibliography}

\end{document}